\documentclass[twocolumn, secnumarabic, amssymb, nobibnotes, aps, showpacs]{revtex4-1}
\usepackage{graphicx,amsmath}
\usepackage{color,multirow,makecell}
\usepackage[bookmarks=false]{hyperref}
\hypersetup{colorlinks=true,citecolor=blue,linkcolor=blue,urlcolor=blue,pdfstartview=FitH,bookmarksopen=true}
\begin{document}

\title{Steady three-dimensional dark state entanglement in dissipative Rydberg atoms}%

\author{S. L. Su}%
\affiliation{School of Physics and Engineering, Zhengzhou University, Zhengzhou 450001, China}
\date{\today}%
\begin{abstract}
Scheme to prepare three-dimensional entangled state between a pair of Rydberg atoms is proposed via dissipative dynamics and Electromagnetic Induced Transparency (EIT) associated with the single-atom dark state. The prepared entangled state is the dark state of the whole system.
The schemes are feasible no matter the system initially in arbitrary purity or mixed states and do not have accurate requirements on evolution time.  In contrast to most of the former Rydberg-atom-based dissipative schemes, the Rydberg-Rydberg interaction~(RRI)~strength do not need to satisfy a certain relation with laser detuning since it works in the blockade as well as intermediate regimes. 
\end{abstract}
\maketitle

\section{introduction}
Rydberg atoms are the neutral atoms with high lying excited state, and they would exhibit strong Rydberg-Rydberg interaction~(RRI)~when close enough~\cite{TF:1994, MTK:2010, Saffman2016}. The most interesting thing caused by RRI is the Rydberg blockade, which has been observed between two Rydberg atoms through sequent~\cite{ett2009} and collective driving~\cite{ayt2009}, respectively.  Besides, the RRI has also been directly measured in experiment~\cite{lar2013}. For quantum information processing~(QIP)~tasks, quantum logic gates and entangled states are basic building block and resource, respectively~\cite{mi2000}. The pioneering works relevant to Rydberg quantum logic gate were proposed by Jaksch \emph{et al}.~\cite{DJP:2000}, in which the model conditions $V~{(\rm RRI~strength)}\gg\Omega~{(\rm Rabi~frequency)}$ and $u\ll\Omega$ are considered, respectively. Lukin \emph{et al}.~\cite{MMR:2001} then describe a method for the
coherent manipulation of quantum information stored in collective excitations of mesoscopic many-Rydberg-atom ensembles. These two works~\cite{DJP:2000,MMR:2001} open a new chapter for Rydberg-atom-based QIP and has been followed by a varity of interesting works,  typically including preparation of quantum entanglement~\cite{dlk2008,mih2009, mk2009}, construction of quantum logic gate~\cite{mt2005,mta2006,lmk2011,hzs2010}, quantum simulators~\cite{hmi2009}, quantum algorithms~\cite{klm2011},  and quantum repeaters~\cite{ybk2010}.

Dissipation induced by the coupling between quantum system and its environment is inevitable and always be considered detrimental for QIP tasks. Commonly, there are three methods to deal with the dissipation: (i) Quantum error correction method~\cite{aaj1996}, which relies on the high-fidelity quantum gate for detecting and correcting errors. (ii) Dynamical decoupling method~\cite{les1999}, which seeks to minimize the unwanted system-bath interactions in an open quantum system but can never completely avoid all unitary errors~\cite{gd2013}.  (iii) Decoherence-free subspace method~\cite{gka1996}, which requires the symmetric coupling between system and its bath.
Different from the above methods that are trying to avoid the influence of dissipation, dissipative dynamics method~\cite{msa1999} opens a new path to deal with decoherence since the dissipation plays a significant role to realize QIP tasks~\cite{saa2008}. 

The dissipative-dynamics-based schemes in Rydberg atoms are interesting since they combine the advantages of Rydberg atom and of dissipative dynamics together. Recently,  Petrosyan and M{\o}lmer found that the atomic spontaneous emission of the intermediate excited state facilitates a single excitation of the Rydberg atom ensemble with nearly unit probability~\cite{dk2013}, which is meaningful for the construction of Rydberg superatom. And Li, Ates, and Lesanovsky studied the dissipative blockade for excited Rydberg atoms~\cite{wci2013}. Then, Rao and M{\o}lmer~\cite{Rao2013}, and Carr and Saffman~\cite{am2013} proposed pioneering works to prepare steady entangled state in Rydberg atoms via the dissipation. The scheme proposed in Ref.~\cite{Rao2013} works under the blockade regime($V\gg\Omega$) as well as intermediate regime ($V\sim\Omega$) based on the Electromagnetic Induced Transparency~(EIT). The steady entangled state contains the dark state induced by the EIT regime. Then the many-body steady entangled state is studied~\cite{Rao2014,sjk2015} based on the RRI. On the other hand, the scheme proposed in Ref.~\cite{am2013} works under the Rydberg antiblockade regime~\cite{ctt2007} which requires the Rydberg pumping condition $V=2\Delta\gg\Omega$ ($\Delta$ denotes laser detuning) and followed by many works~\cite{xjt2014}.

In this manuscript, inspired by the pioneering work in Ref~\cite{Rao2013}, we design a scheme to prepare the three-dimensional  entangled state, which can enhance the security and capacity of QIP~\cite{ajg2011} and violate the local realism more strongly than the two-dimensional entangled state~\cite{dp2000}. The prepared state is the dark state of the whole system which involves the single-atom dark state induced by the dissipative EIT and thus robust on RRI-induced mechanical effect. In addition, the scheme is insensitive to the RRI because it works under the blockade as well as the intermediate regimes and has short convergence time approaches steady state. Finally, we try to translate the dark state entangled state to the ground state entangled state via the adiabatic passage method. 

The rest part of the manuscript is organized as follows. In Sec.~\ref{se2}, the basic model of the scheme, including the dissipative EIT dark state and the approximations based on the dressed state, are introduced. In Sec.~\ref{se3}, the basic dynamics and the performance of the scheme (including fidelity, purity and negativity) are shown based on one group of specific parameters. In Sec.~\ref{se4}, we discuss the robustness of the scheme for a wide range of parameters, try to transfer the dark-state-basis-based entangled state to ground state subspace, and consider experimental feasibility in some ways. The conclusions are given in Sec.~\ref{se5}.

\section{Basic Model}\label{se2}
\begin{figure}
  \centering
  \includegraphics[width=0.8\linewidth]{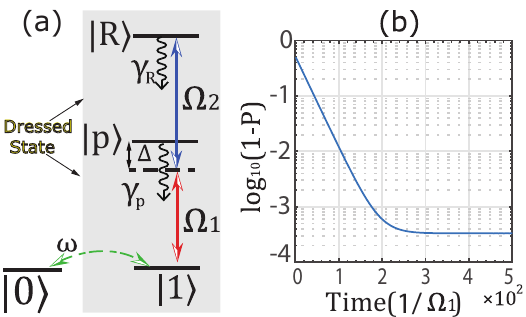}\\
  \caption{(a) Schematic diagram of single Rydberg atom to illustrate the dressed state of single-atom and steady state under dissipation. $|R\rangle$ denotes the long-lived Rydberg state. $|p\rangle$ denotes the short-lived metastable excited state, whose spontaneous emission is crucial and used for preparation of entanglement. $|0\rangle$ and $|1\rangle$ are two ground states. $\Omega_{1}$ and $\Omega_{2}$ denote the Rabi frequencies of the coupling $|1\rangle\leftrightarrow|r\rangle$ and $|r\rangle\leftrightarrow|R\rangle$, and $\Delta$ is the corresponding detuning. $\omega$ is the microwave Rabi frequency or effective Raman coupling strength of the transition $|0\rangle\leftrightarrow|1\rangle$. $\gamma_{R}$ and $\gamma_{p}$ denote the atomic spontaneous emission rates of state $|R\rangle$ and $|p\rangle$, respectively,  where $\gamma_{R}\ll\gamma_{p}$ is satisfied. (b) Population of $|D\rangle$ versus the evolution time. The initial state is $|1\rangle$. Parameters are chosen as $\Omega_{2}=\Omega_{1}$,  $\omega=0$, $\gamma_{p}/\Omega_{1}=0.1515$, $\gamma_{R}/\Omega_{1} = 5\times10^{-5}$. }\label{f1}
\end{figure}


\subsection{Dissipative EIT steady state of single atom}\label{s2.1}
We first consider single Rydberg atom with ladder-type energy level structure, as shown in the shaded area of Fig.~\ref{f1}(a).
The Hamiltonian is 
\begin{equation}\label{eq01}
\hat{H}_{\rm Dr} = \Delta|p\rangle\langle p|+(\Omega_{1}|1\rangle\langle p|+\Omega_{2}|p\rangle\langle R|+{\rm H.c.}).
\end{equation}
It is obvious that the laser fields would induce EIT phenomenon~\cite{ma2005,dk2013} and the system has a dark state $|D\rangle = (\Omega_{2}|1\rangle-\Omega_{1}|R\rangle)/\sqrt{\Omega_{1}^2+\Omega_{2}^2}$. 
Generally speaking, the systematic dynamics could be described well by the master equation 
\begin{equation}\label{eq02}
\dot{\hat{\rho}} = -i[\hat{H},~\hat{\rho}]+\sum_{j}\mathcal{D}[\hat{\sigma}_{j}]\hat{\rho},
\end{equation}
where $\mathcal{D}[\hat{a}]\hat{\rho} \equiv \hat{a}\hat{\rho}\hat{a}^{\dag}-(\hat{\rho}\hat{a}^{\dag}\hat{a}+\hat{a}^{\dag}\hat{a}\hat{\rho})/2$, $\hat{\sigma}_{1}=\sqrt{\gamma_{R}/2}|1\rangle\langle R|$, $\hat{\sigma}_{2}=\sqrt{\gamma_{R}/2}|p\rangle\langle R|$ and $\hat{\sigma}_{3}=\sqrt{\gamma_{p}}|1\rangle\langle p|$.
Fig.~\ref{f1}(b) shows that the steady state of the single Rydberg atom~(without consideration of $|0\rangle$) approaches to the dark state $|D\rangle$ in a short period of evolution time under dissipation.

\subsection{Approximation with the dressed states}\label{s2.2}
To prepare the entangled state, the coupling between ground states are required. We first consider two ground states $|0\rangle$ and $|1\rangle$, as shown in Fig.~\ref{f1}(a) . And the Hamiltonian is
\begin{equation}\label{eq03}
\hat{H}_{\omega} = \omega|0\rangle\langle1|+{\rm H.c.}.
\end{equation}
If $\omega$ is much smaller than $\Omega_{1}$ and $\Omega_{2}$, Hamiltonian~(\ref{eq01}) constitutes dressed states as
\begin{equation}
|D\rangle,~|\zeta\rangle_{\pm} = (2\Omega_{1}|1\rangle +(\Delta\pm\tilde{\Delta})|p\rangle+2\Omega_{2}|R\rangle)/\mathcal{N_{\pm}},
\end{equation}
with the corresponding eigenvalues $E_{0}=0$, $E_{\pm}=(\Delta\pm\tilde{\Delta})/2$, where $\tilde{\Delta}=\sqrt{\Delta^2+4\Omega_{1}^2+4\Omega_{2}^2}$.
Thus, the whole Hamiltonian of single atom ($\hat{H}_{\rm Dr} + \hat{H}_{\omega}$) could be rewritten as
\begin{equation}
\hat{H}_{\omega}' = \omega(\langle 1|D\rangle|0\rangle\langle D|e^{iE_{0}t}+\langle 1|\zeta_{\pm}\rangle|0\rangle\langle \zeta_{\pm}|e^{i E_{\mp}t})+{\rm H.c.}
\end{equation}
after rotating with respect to the dressed states.
If $E_{\pm}\gg E_{0}$, after discarding the high-frequency oscillating terms, one can approximately get 
\begin{equation}
\hat{H}_{\omega}' = \frac{\omega\Omega_{2}}{\sqrt{\Omega_{1}^2+\Omega_{2}^2}}|0\rangle\langle D|+{\rm H.c.}.
\end{equation}
Sec.~\ref{s2.1} shows that the steady state of the dressed state space is the dark state $|D\rangle$, and sec.~\ref{s2.2} shows that replacing the coupling $|0\rangle\leftrightarrow|1\rangle$ with $|0\rangle\leftrightarrow|D\rangle$ is reasonable. In the following subsection we would describe the basic dynamics of the system.

\section{Preparation of three-dimensional entangled dark state}\label{se3}

\subsection{Desired Entangled State}
The three-dimensional entangled state has the form~\cite{ajg2011} 
\begin{equation}\label{3d}
|{\rm \Psi}\rangle=(|00\rangle+|11\rangle+|22\rangle)/\sqrt{3},
\end{equation}
which can enhance the security of QIP~\cite{ajg2011} and violate the local realism more strongly than the two-dimensional entangled state~\cite{dp2000}. In this manuscript, combing with the single-atom EIT dark state, we aim to prepare 
\begin{equation}
|\mathbb{D}\rangle=(|D_{0}\rangle|D_{0}\rangle+|D_{1}\rangle|1\rangle+|2\rangle|D_{2}\rangle)/\sqrt{3},
\end{equation}
where $|D_{j}\rangle=(\Omega_{2}|j\rangle-\Omega_{1}|R_{j}\rangle)/\sqrt{\Omega_{1}^2+\Omega_{2}^2}$ is the dark state of the Hamiltonian $\hat{H}=\Delta|p_{j}\rangle\langle p_{j}|+\Omega_{1}|j\rangle\langle p_{j}|+\Omega_{2}|p_{j}\rangle\langle R_{j}|+{\rm H.c.}$. 

\subsection{Configuration and Hamiltonian}
\begin{figure}
  \centering
  \includegraphics[width=0.9\linewidth]{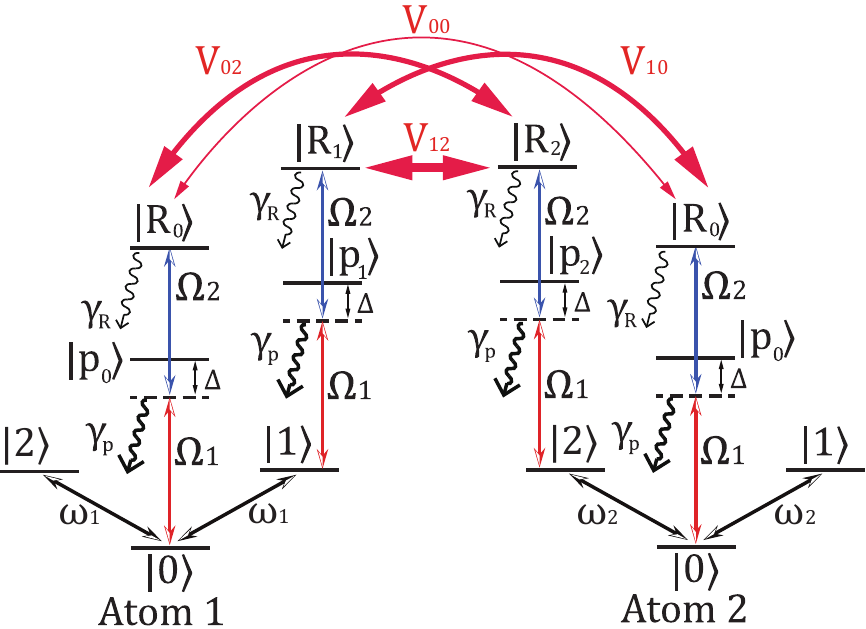}\\
  \caption{Configuration of two Rydberg atoms to prepare three-dimensional entangled state. $|R_{j}\rangle(j=0,1,2)$ denotes the long-lived Rydberg state. $|p_{j}\rangle$ denotes the short-lived metastable excited state, whose spontaneous emission is used for preparation of entanglement. $|j\rangle$ denotes ground state. $\Omega_{1}$ and $\Omega_{2}$ denote the Rabi frequencies of the coupling $|j\rangle\leftrightarrow|p_{j}\rangle$ and $|p_{j}\rangle\leftrightarrow|R_{j}\rangle$, and $\Delta$ is the two-photon detuning. $\omega_{k}$ is the microwave Rabi frequency or effective Raman coupling strength of the transition $|1\rangle\leftrightarrow|0\rangle\leftrightarrow|2\rangle$ of atom \emph{k}. And $\omega_{1}=-\omega_{2}$ is satisfied. $\gamma_{R}$ and $\gamma_{p}$ denote the atomic spontaneous emission rates of $|R\rangle$ and $|p\rangle$, respectively. $V_{mn}$ is the RRI between Rydberg state $|R_{m}\rangle$ and $|R_{n}\rangle$. }\label{f2}
\end{figure}
We consider two Rydberg atoms as shown in Fig.~\ref{f2}. Both of the atoms have two Rydberg states, two metastable states and three ground states. These two atoms interact with each other through the van-der-Waals-type RRI. The Hamiltonian of the whole system can be written as $\hat{H}=\hat{H}_{\Omega}+\hat{H}_{\omega}+\hat{V}$, in which
\begin{eqnarray}
\hat{H}_{\Omega}&=&\sum_{m=0,1}(\Omega_{1}|m\rangle_{1}\langle p_{m}|+\Omega_{2}|p_{m}\rangle_{1}\langle R_{m}|+{\rm H.c.})\cr\cr&&+\sum_{n=0,2}(\Omega_{1}|n\rangle_{2}\langle p_{n}|+\Omega_{2}|p_{n}\rangle_{2}\langle R_{n}|+{\rm H.c.}),\cr\cr
\hat{H}_{\omega}&=&\sum_{k=1,2}[\omega_{k}(|0\rangle_{k}\langle 2|+|0\rangle_{k}\langle 1|)+{\rm H.c.}],
\cr\cr {\rm and}\cr\cr
\hat{H}_{V}&=&\sum_{m=0,1}\sum_{n=0,2}V_{m,n}|R_{m}\rangle_{1}\langle R_{m}|\otimes|R_{n}\rangle_{2}\langle R_{n}|.
\end{eqnarray}

\subsection{Effective Dynamics}
\begin{figure}
  \centering
  \includegraphics[width=\linewidth]{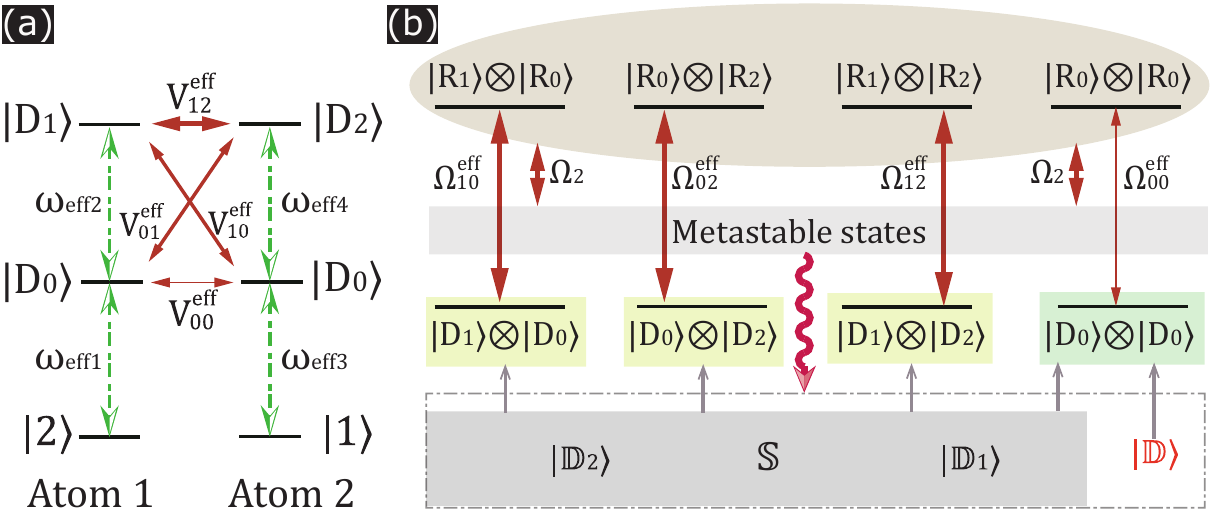}\\
  \caption{(a) Effective coupling processes. $|D_{j}\rangle=(\Omega_{2}|j\rangle-\Omega_{1}|R_{j}\rangle)/\sqrt{\Omega_{1}^2+\Omega_{2}^2}$. (b) Effective unitary and dissipative dynamics. The wavy line denotes dissipative process. The straight lines with two arrows denote effective coupling processes while with single arrows denote that the states in the arrow direction are the components of the states in the negative direction. $|\mathbb{D}\rangle=(|D_{0}\rangle|D_{0}\rangle+|D_{1}\rangle|1\rangle+|2\rangle|D_{2}\rangle)/\sqrt{3}$, $|\mathbb{D}_{1}\rangle=(|D_{0}\rangle|D_{0}\rangle+|D_{1}\rangle|D_{2}\rangle+|2\rangle|1\rangle)/\sqrt{3}$ and $|\mathbb{D}_{2}\rangle=(|D_{0}\rangle|1\rangle+|D_{0}\rangle|D_{2}\rangle+|D_{1}\rangle|D_{0}\rangle+|2\rangle|D_{0}\rangle)/\sqrt{3}$ are three dark states of the effective microwave coupling processes. $\mathbb{S}$ denotes the set of the other six eigenstates of the two-atom effective microwave coupling Hamiltonian. $|\mathbb{D}\rangle$, $|\mathbb{D}_{1}\rangle$ and $|\mathbb{D}_{2}\rangle$ couple with the two-excitation Rydberg states through $\Omega_{\rm eff}$ and further interact with the metastable states through $\Omega_{2}$. The metastable states have short lifetime and would decay to the ground states quickly. Then, if $\Omega_{00}^{\rm eff}\ll\{\Omega_{10}^{\rm eff}, \Omega_{02}^{\rm eff}, \Omega_{12}^{\rm eff} \}$ is satisfied, $|\mathbb{D}\rangle$ would be the steady state of the whole system under the cooperation of unitary and dissipative dynamics.}\label{f3}
\end{figure}
We now analyze the effective dynamics of the whole system. Based on the process similar to that discussed in Sec.~\ref{s2.2}, the Hamiltonian of atom 1 can be approximated to 
\begin{equation}
\hat{H}_{\rm eff1}=\omega_{\rm eff 1}|D_{0}\rangle\langle 2|+\omega_{\rm eff 2}|D_{0}\rangle\langle D_{1}|+{\rm H.c.}
\end{equation}
and of atom 2 can be approximated to
\begin{equation}
\hat{H}_{\rm eff2}=\omega_{\rm eff 3}|D_{0}\rangle\langle 1|+\omega_{\rm eff 4}|D_{0}\rangle\langle D_{2}|+{\rm H.c.},
\end{equation}
as shown in Fig.~\ref{f3}(a). The effective microwave coupling can be calculated through
\begin{equation}
\omega_{1(2)}^{\rm eff} = \langle D_{0}|\hat{H}_{\omega1}|2(D_{1})\rangle,~~\omega_{3(4)}^{\rm eff} = \langle D_{0}|\hat{H}_{\omega2}|1(D_{2})\rangle,~~\end{equation}
where $\hat{H}_{\omega k} = \omega(|1\rangle_{k}\langle0|+|1\rangle_{k}\langle0|)+{\rm H.c.}$ denotes the microwave field Hamiltonian of the \emph{k}-th Rydberg atom. And the effective interaction strength between $|D\rangle_{m}$ of atom 1 and $|D\rangle_{n}$ of atom 2 can be calculated through
\begin{equation}
V_{mn}^{\rm eff} = \langle D_{n}|\langle D_{m}|\hat{H}|D_{m}\rangle|D_{n}\rangle,
\end{equation}
where we define $\langle D_{n}|\langle D_{m}|$ is the conjugate transpose of $|D_{m}\rangle|D_{n}\rangle$ with $m = 0, 1$ and $n = 0, 2$.
Fig.~\ref{f3}(b) describes the effective dynamics of the whole system. The effective coupling between $|D_{m}\rangle|D_{n}\rangle$ and $|R_{m}\rangle|R_{n}\rangle$,  $\Omega_{mn}^{\rm eff}$, can be calculated as
\begin{equation}
\Omega_{mn}^{\rm eff} = \langle R_{n}|\langle R_{m}|\hat{H}|D_{m}\rangle|D_{n}\rangle=\Omega_{1}^2V_{m,n}.
\end{equation}

The effective dynamics of the whole system can be illustrated as follows.
Without consideration of RRIs, the two-atom Hamiltonian has three dark states $|\mathbb{D}\rangle$, $|\mathbb{D}_{1}\rangle$ and $|\mathbb{D}_{2}\rangle$ and six bright eigenstates denoted by the set $\mathbb{S}$.  $|\mathbb{D}_{1}\rangle$, $|\mathbb{D}_{2}\rangle$ and $\mathbb{S}$ contain at least one of the states $|D_{1}\rangle|D_{0}\rangle$, $|D_{0}\rangle|D_{2}\rangle$ and  $|D_{1}\rangle|D_{2}\rangle$ and maybe contain $|D_{0}\rangle|D_{0}\rangle$ among the state $|D_{m}\rangle|D_{n}\rangle$. While $|\mathbb{D}\rangle$ only contain $|D_{0}\rangle|D_{0}\rangle$ among the state $|D_{m}\rangle|D_{n}\rangle$. As mentioned above, the two-atom dark state $|D_{m}\rangle|D_{n}\rangle$ couples to the two-excitation Rydberg state $|R_{m}\rangle|R_{n}\rangle$ via the strength $\Omega_{mn}^{\rm eff}$. And $|R_{m}\rangle|R_{n}\rangle$ couples with the metastable states which would decay to the space $\mathcal{S}\equiv\{|\mathbb{D}\rangle$, $|\mathbb{D}_{1}\rangle$, $|\mathbb{D}_{2}\rangle$, $\mathbb{S}$ \}. 
If $\Omega_{00}^{\rm eff}\ll\{\Omega_{10}^{\rm eff}, \Omega_{02}^{\rm eff}, \Omega_{12}^{\rm eff} \}$ is satisfied, the symmetry of the system is broken because the transition path from $|\mathbb{D}\rangle$ to the two-excitation Rydberg states could be neglected. In this case, given any initial state in the ground state subspace, it could be quickly transformed to the space $\mathcal{S}$. If $|\mathbb{D}\rangle$ is populated, the scheme is succeed. Otherwise the state would be excited and decay, and recycling until $|\mathbb{D}\rangle$ is prepared.
In the next subsection, we would consider the performance of the scheme.

\subsection{Performance of the scheme}
We get the final state of the whole system through numerically solving the master equation
\begin{equation}
\dot{\hat{\rho}}=-i[\hat{H},\hat{\rho}]+\sum_{l=1}^{8}\sum_{j=0}^{2}\big[\hat{\mathcal{L}}_{j}^{l}\rho\hat{\mathcal{L}}_{j}^{l\dag}  -\frac{1}{2}(\hat{\mathcal{L}}_{j}^{l\dag}\hat{\mathcal{L}}_{j}^{l}\rho+\rho\hat{\mathcal{L}}_{j}^{l\dag}\hat{\mathcal{L}}_{j}^{l})  \big],
\end{equation}
where $\hat{\mathcal{L}}_{j}^{1} = \sqrt{\gamma_{R}/3}|j\rangle_{1}\langle R_{0}|,~ \hat{\mathcal{L}}_{j}^{2} = \sqrt{\gamma_{R}/3}|j\rangle_{1}\langle R_{1}|,~\hat{\mathcal{L}}_{j}^{3} = \sqrt{\gamma_{p}/3}|j\rangle_{1}\langle p_{0}|,~\hat{\mathcal{L}}_{j}^{4} = \sqrt{\gamma_{p}/3}|j\rangle_{1}\langle p_{1}|,~\hat{\mathcal{L}}_{j}^{5} = \sqrt{\gamma_{R}/3}|j\rangle_{2}\langle R_{0}|,~\hat{\mathcal{L}}_{j}^{6} = \sqrt{\gamma_{R}/3}|j\rangle_{2}\langle R_{2}|,~\hat{\mathcal{L}}_{j}^{7} = \sqrt{\gamma_{p}/3}|j\rangle_{2}\langle p_{0}|$, and $\hat{\mathcal{L}}_{j}^{8} = \sqrt{\gamma_{p}/3}|j\rangle_{2}\langle p_{2}|$ denote the dissipative terms.
\begin{figure}
  \centering
  \includegraphics[width=\linewidth]{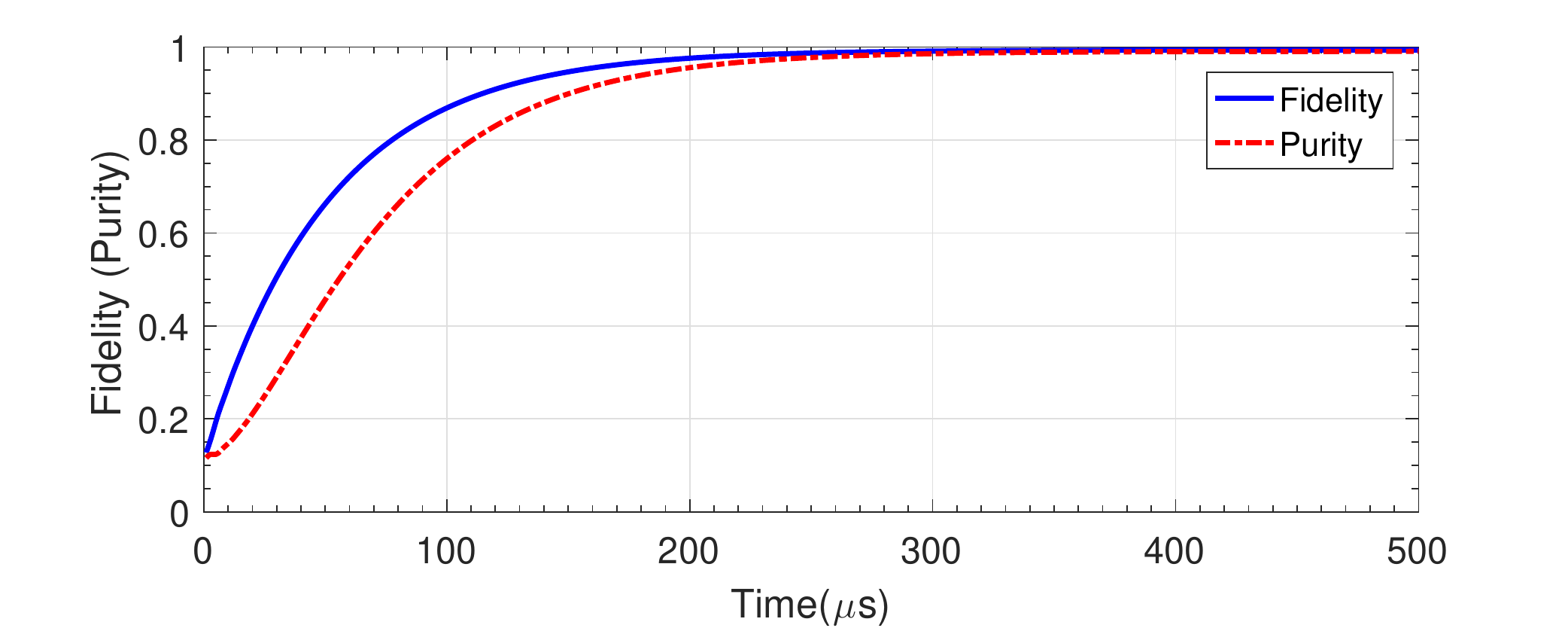}\\
  \caption{Fidelity to prepare $|\mathbb{D}\rangle$ and purity of the system versus evolution time. The parameters are chosen as $\Omega_{1}/2\pi = 30$ MHz, $\Omega_{2}=3.85\Omega_{1}$, $V_{12} = 2\Omega_{1}$, $V_{10}=V_{02}=0.8V_{12}$, $V_{00}=0.001V_{12}$, $\gamma_{p}/2\pi=10$ MHz, $\gamma_{R} = 1$ KHz, $\omega_{1} = 0.004\Omega_{1}$, and $\Delta=0$. Suppose the system is initially in the mixed state $(\sum_{j=0,1,2}\sum_{j'=0,1,2}|j\rangle\langle j|\otimes|j'\rangle\langle j'|)/9$.}\label{f4}
\end{figure}
\begin{figure}
  \centering
  \includegraphics[width=\linewidth]{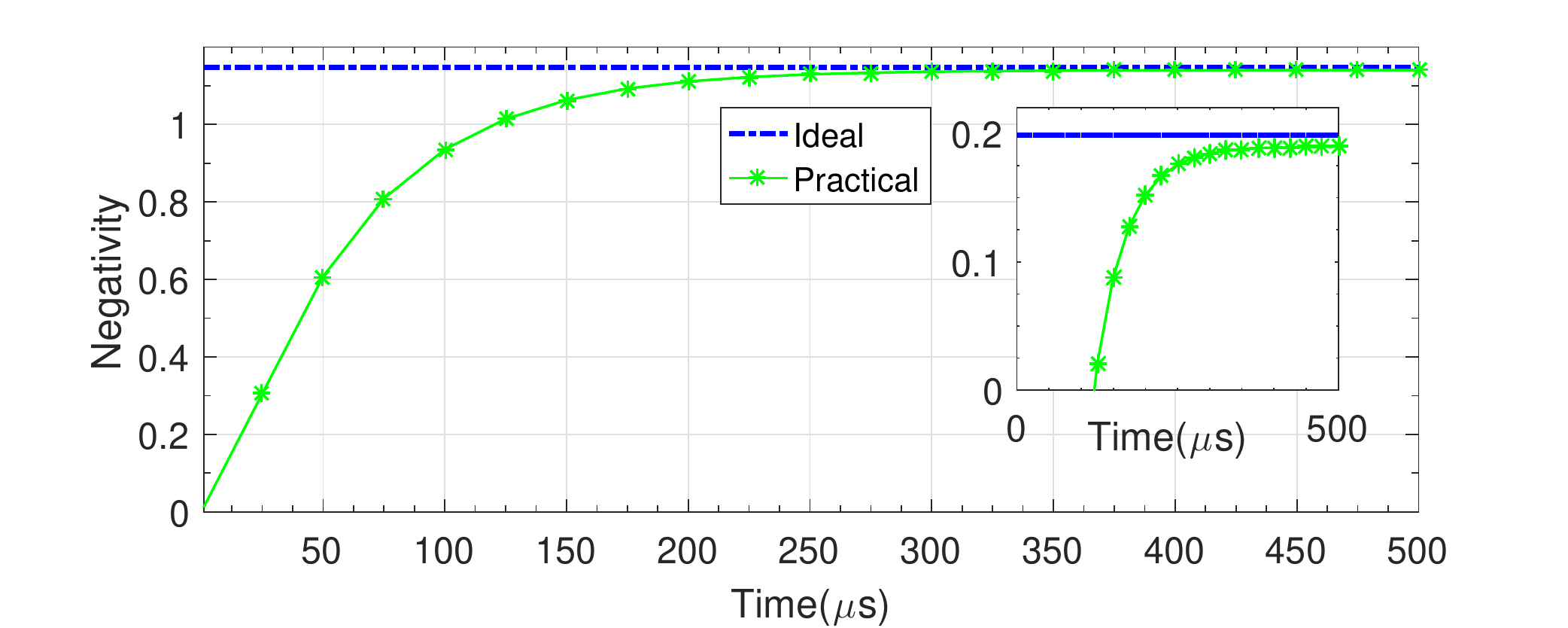}\\
  \caption{Negativity of practical systematic state versus evolution time and the negativity of the ideal state. The inset displays the variation of the logarithmic negativity. The parameters are the same as that in Fig.~\ref{f4}. And the bare state basis are used for calculation.}\label{f5}
\end{figure}

Fidelity denotes the closeness of states in the natural geometry of Hilbert space, which can be calculated through~\cite{r1994} 
\begin{equation}
F(|\mathbb{D}\rangle\langle \mathbb{D}|, \hat{\rho}(t))\equiv\langle\mathbb{D}|\hat{\rho}(t)|\mathbb{D}\rangle.
\end{equation}
Purity is a measure of how pure a quantum state is and can be calculated as~\cite{g2007} 
\begin{equation}
P(\hat{\rho}(t))\equiv{\rm Tr}[\hat{\rho}(t)^2].
\end{equation}
One can conclude that the scheme performs well when both of fidelity and purity are very close to 1. In addition, negativity, a measure deriving from the PPT criterion for separability~\cite{kpa1998}, has shown to be a proper measure of entanglement with the definition~\cite{kpa1998}
\begin{equation}
\mathcal{N}(\hat{\rho}) =\frac{{\rm Tr}\sqrt{(\rho^{T_{A}})^{\dag}(\rho^{T_{A}})} -1 }{2}.
\end{equation}
Then, the logarithmic negativity is proposed with the definition~\cite{m2005}
\begin{equation}
E_{\mathcal{N}}(\hat{\rho}) \equiv {\rm log}_{2}{\rm Tr}\sqrt{(\rho^{T_{A}})^{\dag}(\rho^{T_{A}})}:={\rm log}_{2}(2\mathcal{N}+1).
\end{equation}
Based on the final state $\hat{\rho}(t)$, one can calculate the fidelity, purity and negativity of our scheme with the above definitions, as shown in Figs.~\ref{f4} and \ref{f5} with one group of specific parameters. The results show that both of the fidelity and purity are close to 1 and the negativity is close to the ideal value.

\section{Discussions}\label{se4}

\subsection{Variation of parameters}
Experimentally, the obtainable parameters are not unique. Thus, it is necessary to see the performance of the scheme under various parameters. 
\begin{figure}
  \centering
  \includegraphics[width=0.8\linewidth]{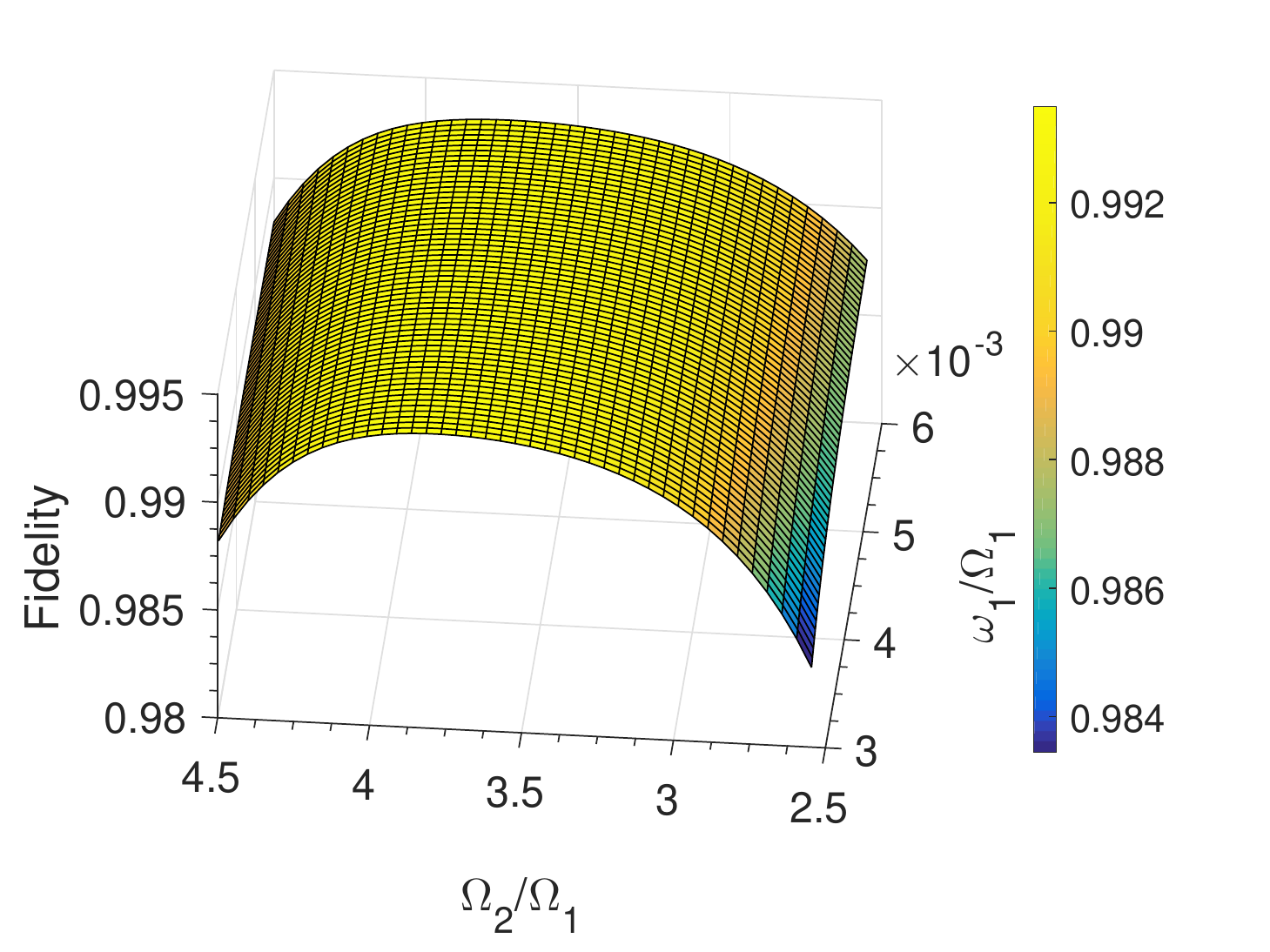}\\
  \caption{Fidelity with respect to $\Omega_{2}/\Omega_{1}$ and $\omega/\Omega_{1}$ at the time $T=500\mu s$. $\Omega_{1}/2\pi = 30$ MHz, $V_{10}=V_{02}=0.5V_{12}$. And the rest parameters are the same as that in Fig.~\ref{f4}.}\label{f6}
\end{figure}
\begin{figure}
  \centering
  \includegraphics[width=\linewidth]{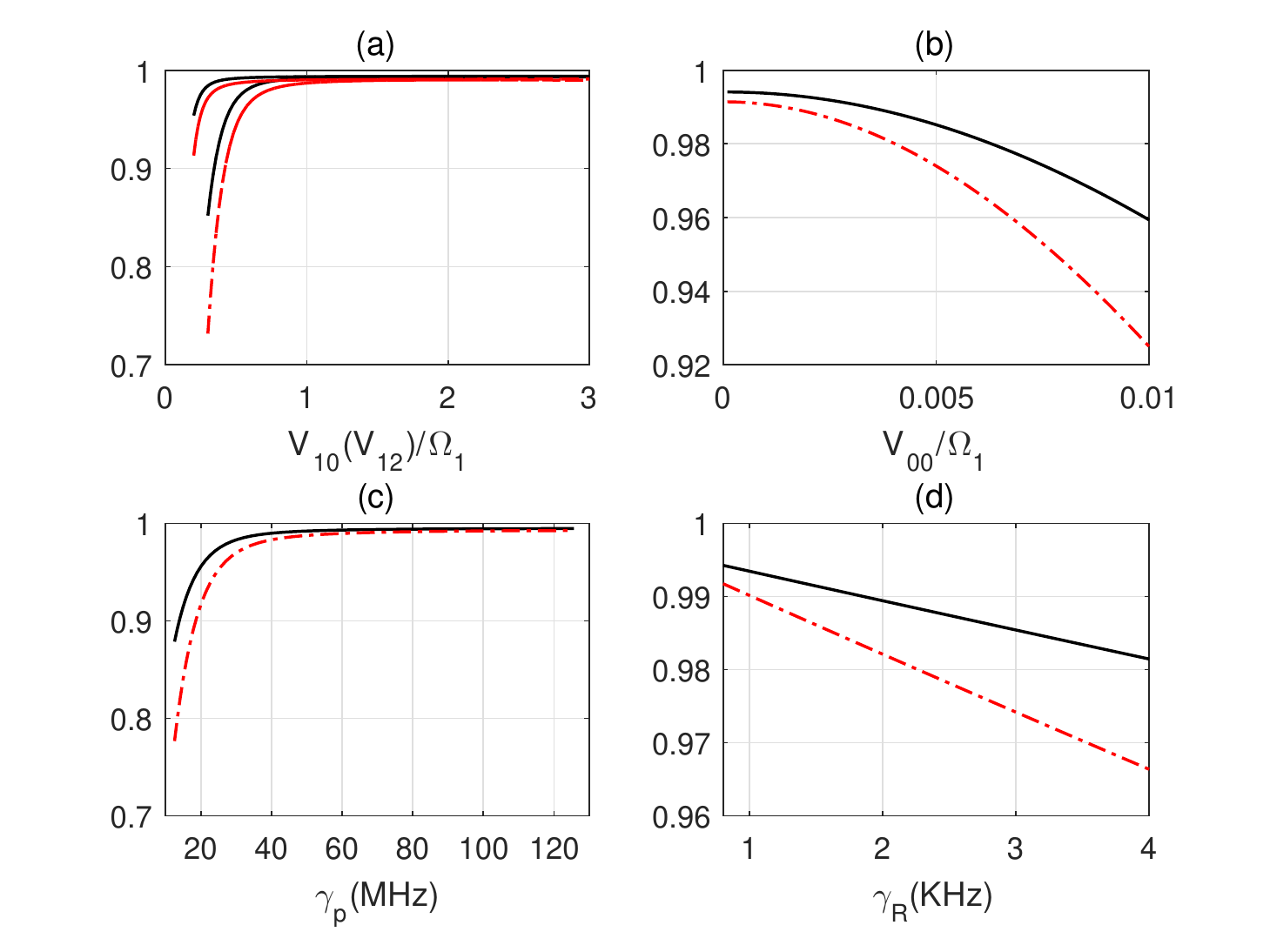}\\
  \caption{Fidelity and purity with respect to different parameters at the time $T=500\mu s$. (a) (From top to bottom) Fidelity versus $V_{10}/\Omega_{1}$; Purity versus $V_{10}/\Omega_{1}$; Fidelity versus $V_{12}/\Omega_{1}$; Purity versus $V_{12}/\Omega_{1}$. (b) Fidelity~(top, solid line) and purity~(bottom, dashed-solid line) versus $V_{00}/\Omega_{1}$. (c) Fidelity~(top, solid line) and purity~(bottom, dashed-solid line) versus $\gamma_{p}$. (d) Fidelity~(top, solid line) and purity~(bottom, dashed-solid line) versus $\gamma_{R}$. The rest parameters are the same as that in Fig.~\ref{f4} and $V_{10}=V_{02}$ is used for simulation.}\label{f7}
\end{figure}
In Fig.~\ref{f6}, we plot the fidelity versus the variations of $\Omega_{2}/\Omega_{1}$ and $\omega_{1}/\Omega_{1}$, which shows the scheme may have a good performance in a wide range values of parameters $\Omega_{2}$ and $\omega_{1}$. In Fig.~\ref{f7}, we plot the fidelity and purity versus $V_{10},~V_{12},~V_{00},~\gamma_{p}$ and $\gamma_{R}$, respectively. The results show a good robustness on the variation of RRI, which is absolutely different to the Rydberg-antiblockade-based schemes and may release the experimental requirements.

\subsection{Transforming the entanglement to ground state subspace}
\begin{figure}
  \centering
  \includegraphics[width=\linewidth]{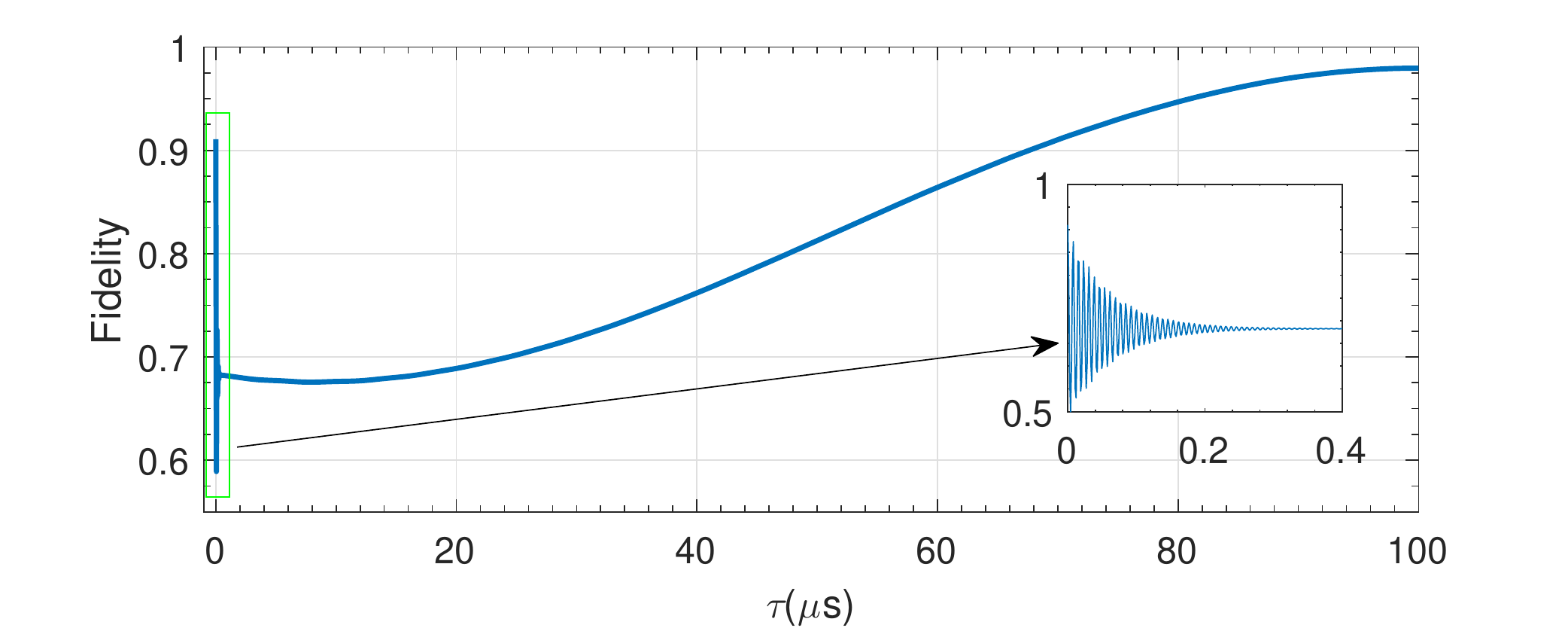}\\
  \caption{Fidelity of the state $|\Psi\rangle$ versus $\tau$. Except $\Omega_{1}$, the other parameters are the same as that in Fig.~\ref{f4}.}\label{f8}
\end{figure}

In this subsection, we try to transfer $|\mathbb{D}\rangle$ to the three-dimensional entangled state in the ground state subspace, as shown in Eq.~(\ref{3d}), via the adiabatic technique. To do this, we should turn off $\Omega_{1}$ adiabatically. For simplicity, we choose $\Omega_{1}(\tau)=\cos[\pi \tau/(2T)]$, and the result is shown in Fig.~\ref{f8}, where the beginning of $\tau$ is the end of the time in Fig.~\ref{f4}. Although the fidelity is not higher than 0.99, the performance could be further improved through using the shortcut to adiabaticity method and design the pulse shape more carefully. And we just numerically show that to transfer the entangled state to the ground state subspace is feasible. 
 
\subsection{Experimental considerations}
For practical experiments, the ground state energy level of the two atoms could be chosen as $|0\rangle \equiv |F=1, m_{f}=0\rangle$, $|1\rangle \equiv |F=1, m_{f}=-1\rangle$ and $|2\rangle \equiv |F=1, m_{f}=+1\rangle$ of 5$S_{1/2}$. And the metastable energy level of the two atoms could be chosen as $|p_{0}\rangle \equiv |F=1, m_{f}=0\rangle$, $|p_{1}\rangle \equiv |F=1, m_{f}=-1\rangle$ and $|p_{2}\rangle \equiv |F=1, m_{f}=+1\rangle$ of 5$P_{3/2}$. Rydberg state energy level should be chosen carefully because the asymmetric RRI and dipole transition selection rules. 

The asymmetric RRI is crucial for our scheme and has been widely used for many Rydberg-atom-based quantum information processing schemes~\cite{mk2009,am2013,eak2007,lmk2011}. In Ref.~\cite{mk2009}, some groups of reasonable asymmetric RRI parameters are used and predicted to satisfy $\{\Delta_{sp}, \Delta_{ss}\}\gg\Delta_{pp}$, where $\Delta_{sp}$, $\Delta_{ss}$ and $\Delta_{pp}$ correspond to $V_{10}(V_{02})$, $V_{12}$ and $V_{00}$ of the present scheme, respectively. The Rydberg states considered in Ref.~\cite{mk2009} are $|s\rangle=|41s_{1/2},~m=1/2\rangle$ and $|p\rangle=|40p_{3/2},~m=1/2\rangle$, and the maximal, minimum and average asymmetry \big[$\Delta_{sp}(\sim n^4/R^3)/\Delta_{pp}(\sim n^{11}/R^6)$\big] are about 1400, 185, 757, respectively. However, we cannot use these results directly because the lifetime $\tau$ of $n=40$ is about $57\mu s$ which is not enough for our scheme to achieve a high fidelity. From this point, Rydberg states with higher principal quantum numbers are preferred because $\tau\sim n^3$~\cite{TF:1994,MTK:2010}. Nevertheless, for a fixed distance, the principal quantum number should not be too large because the asymmetry $\sim n^{-7}$. Based on these rules, if we choose $|R_{0}\rangle\equiv|70p_{3/2}~m=1/2\rangle$(for two atoms), $|R_{1}\rangle\equiv|71s_{1/2}~m=3/2\rangle$(for atom 1) and $|R_{2}\rangle\equiv|71s_{1/2}~m=3/2\rangle$(for atom 2), the maximal asymmetry about 27.8523 and the lifetime $\tau\simeq305\mu s$ are achieved. It should be noted that the transition between $|p_{0}\rangle$ and $|R_{0}\rangle$ could be realized via introducing a large detuning interaction with intermediate $|s\rangle$ or $|d\rangle$ state(similar to the degenerate two-photon process). With these assumptions and assume the other factors which influence the RRI keep invariant, the fidelity and purity are estimated about 0.6047 and 0.507, respectively, at $\tau\simeq305\mu s$ with $\Omega_{1}$, $\Omega_{2}$ and $\omega$ parameters the same as that in Fig.~\ref{f4}. Similarly, we roughly estimate the performance for different principal quantum  numbers at the corresponding lifetime with the maximal asymmetry for different principal quantum numbers, as shown in Fig.~\ref{f7}.
\begin{figure}
  \centering
  \includegraphics[width=\linewidth]{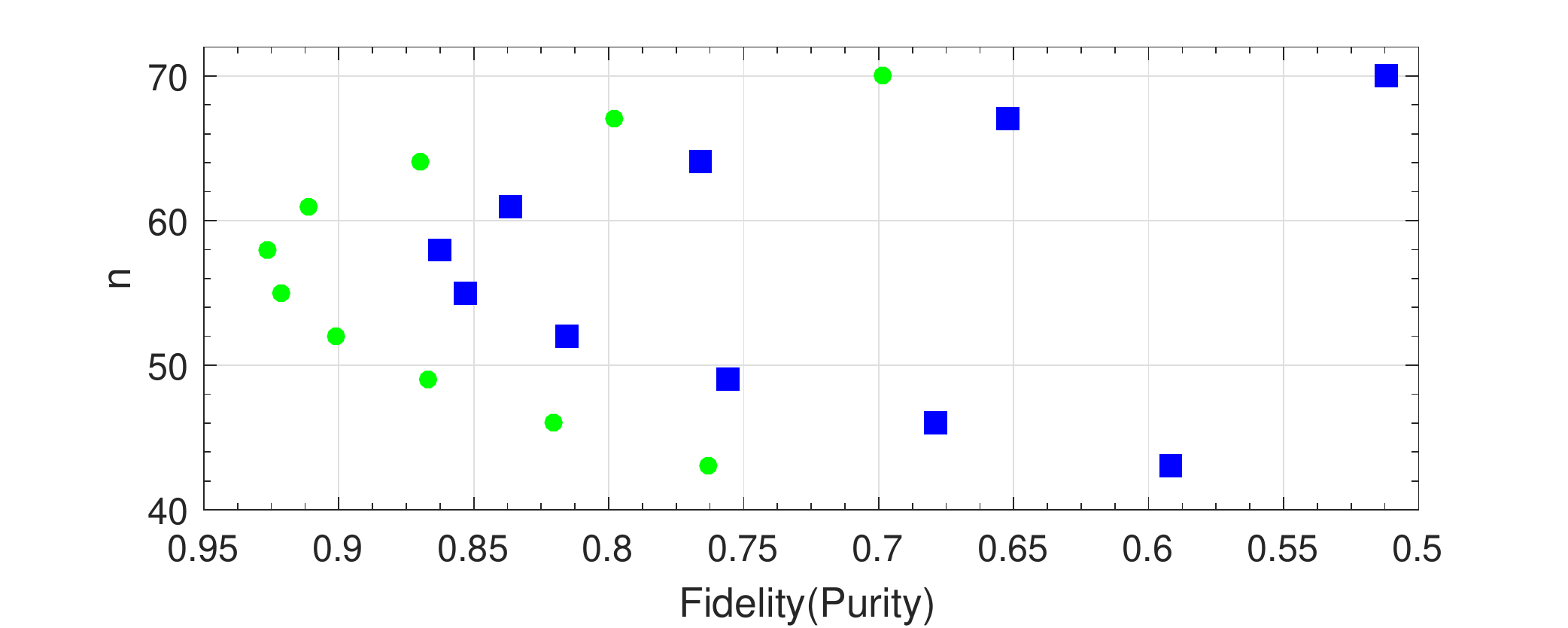}\\
  \caption{Fidelity and purity versus principal quantum number \emph{n} with assumptions discussed in the text. Green circle denotes the fidelity and the blue square denotes the purity. We assume $\gamma_{p}/(2\pi)=10$~MHz and consider variable $\gamma_{R}$ for different \emph{n}.}\label{f9}
\end{figure}
 We use the dipole-dipole interactions rather than vdW interaction for two different Rydberg states for simulation. 
 
Besides, we could also choose $|R_{0}\rangle\equiv|75p_{3/2}~m=1/2\rangle$(for two atoms), $|R_{1}\rangle\equiv|125s_{1/2}~m=3/2\rangle$(for atom 1) and $|R_{2}\rangle\equiv|120s_{1/2}~m=3/2\rangle$(for atom 2). We suppose the interaction between states $|R_{0}\rangle_{1}$ and $|R_{0}\rangle_{2}$~($|R_{1}\rangle_{1}$ and $|R_{2}\rangle_{2}$) is van der Waals (vdW) interaction with strength $V_{00}$~($V_{12}$) while between $|R_{1}\rangle_{1}$ and $|R_{0}\rangle_{2}$~($|R_{0}\rangle_{1}$ and $|R_{2}\rangle_{2}$) is dipole-dipole interaction with strength $D_{10}$~($D_{02}$). Due to the principal quantum number differences, vdW strength $V_{12}$ could be larger than $V_{00}$. Meanwhile, the Stark tuned F\"{o}rster resonances have been demonstrated experimentally~\cite{idi2010,shd2014} due to the dipole-dipole interaction and the detuning could be adjusted through the electric field. Then, one could achieve the non-resonant dipole-dipole interaction induced vdW between $|R_{1}\rangle_{1}$ and $|R_{0}\rangle_{2}$~($|R_{0}\rangle_{1}$ and $|R_{2}\rangle_{2}$). Under this case, the higher fidelity and purity may be achieved due to the long lifetime Rydberg states.

\section{Conclusion}\label{se5}
In this manuscript, we proposed a scheme to prepare three-dimensional dark state entanglement based on the dissipative EIT and Rydberg blockade regimes. In contrast to the Rydberg-antiblockade-based schemes, the present one is insensitive to the variation of RRI and has a shorter evolution time approaches steady state, which may release the experimental requirements. In addition, the scheme does not have accurate requirement on the evolution time and is feasible even the system is initially in the mixed state.

\begin{center}
{\bf{ACKNOWLEDGMENT}}
\end{center}
This work was supported by National Natural Science Foundation of China under Grant No. 11804308.


\begin{thebibliography}{999}

\bibitem{TF:1994} T. F. Gallagher, Rydberg Atoms (Cambridge University Press, Cambridge, UK, 1994).

\bibitem{MTK:2010} M. Saffman, T. G. Walker, and K. M{\o}lmer, \href{https://doi.org/10.1103/RevModPhys.82.2313}{Rev. Mod. Phys. \textbf{82}, 2313 (2010)}. 

\bibitem {Saffman2016} M. Saffman, \href{https://doi.org/10.1088/0953-4075/49/20/202001}  {J. Phys. B At. Mol. Opt. Phys. \textbf{49}, 202001 (2016)}.

\bibitem{ett2009} E. Urban, T. A. Johnson, T. Henage, L. Isenhower, D. D. Yavuz, T. G.Walker, and M. Saffman, \href{https://www.nature.com/articles/nphys1178} {Nat. Phys. \textbf{5}, 110 (2009).}

\bibitem{ayt2009} A. Ga\"{e}tan, Y. Miroshnychenko, T. Wilk, A. Chotia, M. Viteau, D. Comparat, P. Pillet, A. Browaeys, and P. Grangier, \href{https://www.nature.com/articles/nphys1183} {Nat. Phys. \textbf{5}, 115 (2009).}

\bibitem{lar2013} L. B\'eguin, A. Vernier, R. Chicireanu, T. Lahaye, and A. Browaeys, \href{https://journals.aps.org/prl/abstract/10.1103/PhysRevLett.110.263201} {Phys. Rev. Lett. \textbf{110}, 263201 (2013)}.

\bibitem{mi2000} M. A. Nielsen and I. L. Chuang, \emph{Quantum Computation and Quantum Information}, Cambridge University Press, Cambridge, 2000.

\bibitem{DJP:2000} D. Jaksch, J. I. Cirac, P. Zoller, S. L. Rolston, R. C\^{o}t\'{e}, and M. D. Lukin, \href{https://journals.aps.org/prl/abstract/10.1103/PhysRevLett.85.2208} {Phys. Rev. Lett. \textbf{85}, 2208 (2000)}.

\bibitem{MMR:2001} M. D. Lukin, M. Fleischhauer, R. C\^{o}t\'{e}, L. M. Duan, D. Jaksch, J. I. Cirac, and P. Zoller, \href{https://journals.aps.org/prl/abstract/10.1103/PhysRevLett.87.037901} {Phys. Rev. Lett. \textbf{87}, 037901 (2001).}

\bibitem{dlk2008} D. M{\o}ller, L. B. Madsen, and K. M{\o}lmer, \href{https://journals.aps.org/prl/abstract/10.1103/PhysRevLett.100.170504}{Phys. Rev. Lett. \textbf{100}, 170504 (2008).} 

\bibitem{mih2009} M. M\"{u}ller, I. Lesanovsky, H. Weimer, H. P. B\"{u}chler, and P. Zoller, \href{https://journals.aps.org/prl/abstract/10.1103/PhysRevLett.102.170502}  {Phys. Rev. Lett. \textbf{102}, 170502 (2009).}

\bibitem{mk2009} M. Saffman and K. M{\o}lmer, \href{10.1103/PhysRevLett.102.240502}{Phys. Rev. Lett. \textbf{102}, 240502 (2009).}

\bibitem{mt2005} M. Saffman and T. G. Walker, \href{https://journals.aps.org/pra/abstract/10.1103/PhysRevA.72.022347}{Phys. Rev. A \textbf{72}, 042302 (2005).}

\bibitem{mta2006} M. Cozzini, T. Calarco, A. Recati, and P. Zoller, \href{https://www.sciencedirect.com/science/article/abs/pii/S0030401806004998?via%3Dihub}{Opt. Commun. \textbf{264}, 375 (2006).}

\bibitem{lmk2011} L. Isenhower, M. Saffman, and K. M{\o}lmer, \href{https://link.springer.com/article/10.1007%2Fs11128-011-0292-4}  {Quantum Inf. Process. \textbf{10}, 755 (2011).}

\bibitem{hzs2010} H. Z. Wu, Z. B. Yang, and S. B. Zheng, \href{https://journals.aps.org/pra/abstract/10.1103/PhysRevA.82.034307 } {Phys. Rev. A \textbf{82}, 034307 (2010).}

\bibitem{hmi2009} H. Weimer, M. M\"{u}ller, I. Lesanovsky, P. Zoller, and H. P. B\"{u}chler, \href{https://www.nature.com/articles/nphys1614}{Nat. Phys. \textbf{6}, 382 (2010)}; H. Weimer, M. M\"{u}ller, H. P. B\"{u}chler, and I. Lesanovsky, \href{https://link.springer.com/article/10.1007%2Fs11128-011-0303-5}{Quantum Inf. Process. \textbf{10}, 885 (2011).}

\bibitem{klm2011} K. M{\o}lmer, L. Isenhower, and M. Saffman, \href{http://iopscience.iop.org/article/10.1088/0953-4075/44/18/184016/meta}{J. Phys. B \textbf{44}, 184016 (2011);} A. Chen, \href{https://www.osapublishing.org/oe/abstract.cfm?uri=oe-19-3-2037}{Opt. Express \textbf{19}, 2037 (2011).}

\bibitem{ybk2010} Y. Han, B. He, K. Heshami, C.-Z. Li, and C. Simon, \href{https://journals.aps.org/pra/abstract/10.1103/PhysRevA.81.052311}{Phys. Rev. A \textbf{81}, 052311 (2010);} B. Zhao, M. M\"{u}ller, K. Hammerer, and P. Zoller, \href{https://journals.aps.org/pra/abstract/10.1103/PhysRevA.81.052329}{Phys. Rev. A \textbf{81}, 052329 (2010).}

\bibitem{aaj1996} A. Steane, \href{https://royalsocietypublishing.org/doi/abs/10.1098/rspa.1996.0136} {Proc. R. Soc. A \textbf{452}, 2551–2577 (1996)}; A. R. Calderbank and P. W. Shor, \href{https://journals.aps.org/pra/abstract/10.1103/PhysRevA.54.1098} {Phys. Rev. A \textbf{54}, 1098 (1996)}; J. Chiaverini, et al. \href{https://www.nature.com/articles/nature03074} {Nature (London) \textbf{432}, 602 (2004).}

\bibitem{les1999} L. Viola, E. Knill, and S. Lloyd, \href{https://journals.aps.org/prl/abstract/10.1103/PhysRevLett.82.2417} {Phys. Rev. Lett., \textbf{82} 2417 (1999).}

\bibitem{gd2013} G. A. Paz-Silva and D. A. Lidar. \href{https://www.nature.com/articles/srep01530} {Sci. Rep., \textbf{3},1530 (2013).}

\bibitem{gka1996} G. M. Palma, K. A. Suominen, and A. K. Ekert, \href{https://royalsocietypublishing.org/doi/abs/10.1098/rspa.1996.0029} {Proc. R. Soc. A \textbf{452}, 567–584 (1996)}; L. M. Duan and G. C. Guo, \href{https://journals.aps.org/prl/abstract/10.1103/PhysRevLett.79.1953} {Phys. Rev. Lett. \textbf{79}, 1953–1956 (1997)}; P. Zanardi and M. Rasetti, \href{https://journals.aps.org/prl/abstract/10.1103/PhysRevLett.79.3306} {Phys. Rev. Lett. \textbf{79}, 3306–3309 (1997)}; D. A. Lidar, I. L. Chuang, and K. B. Whaley, \href{https://journals.aps.org/prl/abstract/10.1103/PhysRevLett.81.2594} {Phys. Rev. Lett. \textbf{81}, 2594 (1998).}

\bibitem{msa1999} M. B. Plenio, S. F. Huelga, A. Beige, and P. L. Knight, \href{https://journals.aps.org/pra/abstract/10.1103/PhysRevA.59.2468} {Phys. Rev. A \textbf{59}, 2468–2475 (1999)}; C. Cabrillo, J. I. Cirac, P. Garc\'{\i}a-Fern\'{a}ndez, and P. Zoller, \href{https://journals.aps.org/pra/abstract/10.1103/PhysRevA.59.1025} {Phys. Rev. A \textbf{59}, 1025 (1999).}

\bibitem{saa2008} S. Diehl, A. Micheli, A. Kantian, B. Kraus, H. P. B\"{u}chler, and P. Zoller, \href{https://www.nature.com/articles/nphys1073} {Nat. Phys. \textbf{4}, 878 (2008)}; F. Verstraete, M. M. Wolf, and J. Ignacio Cirac, \href{https://www.nature.com/articles/nphys1342} {Nat. Phys. \textbf{5}, 633 (2009)}; G. Vacanti and A. Beige, \href{http://iopscience.iop.org/article/10.1088/1367-2630/11/8/083008/meta} {New J. Phys. \textbf{11}, 083008 (2009)}; M. J. Kastoryano, F. Reiter, and A. S. S{\o}rensen, \href{https://journals.aps.org/prl/abstract/10.1103/PhysRevLett.106.090502} {Phys. Rev. Lett. \textbf{106}, 090502 (2011)}; W. Qin, A. Miranowicz, P. B. Li, X. Y. L\"{u}, J. Q. You, and F. Nori, \href{https://journals.aps.org/prl/abstract/10.1103/PhysRevLett.120.093601} {Phys. Rev. Lett. \textbf{120} 093601 (2018).} 

\bibitem{dk2013} D. Petrosyan and K. M{\o}lmer, \href{https://journals.aps.org/pra/abstract/10.1103/PhysRevA.87.033416} {Phys. Rev. A \textbf{87}, 033416 (2013).}

\bibitem{wci2013} W. Li, C. Ates, and I. Lesanovsky, \href{https://journals.aps.org/prl/abstract/10.1103/PhysRevLett.110.213005} {Phys. Rev. Lett. \textbf{110}, 213005 (2013).}

\bibitem{Rao2013} D. D. B. Rao and K. M{\o}lmer, \href{https://journals.aps.org/prl/abstract/10.1103/PhysRevLett.111.033606}{Phys. Rev. Lett. \textbf{111}, 033606 (2013)}.

\bibitem{am2013} A. W. Carr and M. Saffman, \href{https://journals.aps.org/prl/abstract/10.1103/PhysRevLett.111.033607} {Phys. Rev. Lett. \textbf{111}, 033607 (2013).}

\bibitem{Rao2014} D. D. B. Rao and K. M{\o}lmer, \href{https://journals.aps.org/pra/abstract/10.1103/PhysRevA.90.062319}{Phys. Rev. A \textbf{90}, 062319 (2014)}.

\bibitem{sjk2015} S. K. Lee, J. Cho and K. S. Choi, \href{http://iopscience.iop.org/article/10.1088/1367-2630/17/11/113053/meta} {New J. Phys. \textbf{17} 113053 (2015).}

\bibitem{ctt2007} C. Ates, T. Pohl, T. Pattard, and J. M. Rost, \href{https://journals.aps.org/prl/abstract/10.1103/PhysRevLett.98.023002} {Phys. Rev. Lett. \textbf{98}, 023002 (2007)}; T. Pohl and P. R. Berman, \href{https://journals.aps.org/prl/abstract/10.1103/PhysRevLett.102.013004} {Phys. Rev. Lett. \textbf{102}, 013004 (2009)}; J. Qian, Y. Qian, M. Ke, X. L. Feng, C. H. Oh, and Y. Z. Wang, \href{https://journals.aps.org/pra/abstract/10.1103/PhysRevA.80.053413} {Phys. Rev. A \textbf{80}, 053413 (2009)}; T. Amthor, C. Giese, C. S. Hofmann, and M. Weidem\"{u}ller, \href{https://journals.aps.org/prl/abstract/10.1103/PhysRevLett.104.013001} {Phys. Rev. Lett. \textbf{104}, 013001 (2010)}; Z. C. Zuo and K. Nakagawa, \href{https://journals.aps.org/pra/abstract/10.1103/PhysRevA.82.062328} {Phys. Rev. A \textbf{82}, 062328 (2010)}; T. E. Lee, H. H\"{a}ffner, and M. C. Cross, \href{https://journals.aps.org/prl/abstract/10.1103/PhysRevLett.108.023602} {Phys. Rev. Lett. \textbf{108}, 023602 (2012);} S. L. Su,  Y.  Gao, E. Liang, and S. Zhang, \href{https://journals.aps.org/pra/abstract/10.1103/PhysRevA.95.022319} {Phys. Rev. A, \textbf{95}, 022319 (2017).}

\bibitem{xjt2014} X. Q. Shao, J. Bin You, T. Y. Zheng, C. H. Oh, and S. Zhang, \href{https://journals.aps.org/pra/abstract/10.1103/PhysRevA.89.052313}{Phys. Rev. A \textbf{89}, 052313 (2014)}; S. L. Su, Q. Guo, H. F. Wang, and S. Zhang, \href{https://journals.aps.org/pra/abstract/10.1103/PhysRevA.92.022328}{Phys. Rev. A \textbf{92}, 022328 (2015)}; S. L. Su, Y. Tian, H. Z. Shen, H. Zang, E. Liang, and S. Zhang, \href{https://journals.aps.org/pra/abstract/10.1103/PhysRevA.96.042335}{Phys. Rev. A \textbf{96}, 042335 (2017)}; X. Q. Shao, J. H. Wu, and X. X. Yi, \href{https://journals.aps.org/pra/abstract/10.1103/PhysRevA.95.062339}{Phys. Rev. A \textbf{95}, 062339 (2017)}; X. Chen, H. Xie, G. W. Lin, X. Shang, M. Y. Ye, and X. M. Lin, \href{https://journals.aps.org/pra/abstract/10.1103/PhysRevA.96.042308}{Phys. Rev. A \textbf{96}, 042308 (2017)}; X. Q. Shao, J. H. Wu, and X. X. Yi, \href{https://journals.aps.org/pra/abstract/10.1103/PhysRevA.95.022317}{Phys. Rev. A \textbf{95}, 022317 (2017)} X. Q. Shao, J. H. Wu, X. X. Yi, and G. L. Long, \href{https://journals.aps.org/pra/abstract/10.1103/PhysRevA.96.062315}{Phys. Rev. A \textbf{96}, 062315 (2017)}.

\bibitem{ajg2011} A. C. Dada, J. Leach, G. S. Buller, M. J. Padgett, and E. Andersson, \href{http://www.nature.com/articles/nphys1996}{Nat. Phys. \textbf{7}, 677 (2011)};
~M. Bourennane, A. Karlsson, G. Bj\"{o}rk, N. Gisin, N, and N. J. Cerf, \href{http://iopscience.iop.org/article/10.1088/0305-4470/35/47/307/meta}{J. Phys. A, \textbf{35}, 10065 (2002)};~T. Durt, N. J. Cerf, N. Gisin, and M. \.{Z}ukowski, \href{https://journals.aps.org/pra/abstract/10.1103/PhysRevA.67.012311}{Phys. Rev. A, \textbf{67}, 012311 (2003)};~T. Durt, D. Kaszlikowski, J.-L. Chen, and L. C. Kwek, \href{https://journals.aps.org/pra/abstract/10.1103/PhysRevA.69.032313}{Phys. Rev. A \textbf{69}, 032313 (2004)};~D. Bru{\ss} and C. Macchiavello, \href{https://journals.aps.org/prl/abstract/10.1103/PhysRevLett.88.127901}{Phys. Rev. Lett., \textbf{88}, 127901 (2002)};~N. J. Cerf, M. Bourennane, A. Karlsson, and N. Gisin, \href{https://journals.aps.org/prl/abstract/10.1103/PhysRevLett.88.127902}{Phys. Rev. Lett., \textbf{88}, 127902 (2002)}.

\bibitem{dp2000}D. Kaszlikowski, P. Gnaci{\'n}ski, M. {\.Z}ukowski, W. Miklaszewski, and A. Zeilinger, \href{https://journals.aps.org/prl/abstract/10.1103/PhysRevLett.85.4418}{Phys. Rev. Lett., \textbf{85}, 4418 (2000)}.


\bibitem{ma2005} M. Fleischhauer, A. Imamoglu, and J. P. Marangos, \href{https://journals.aps.org/rmp/abstract/10.1103/RevModPhys.77.633}{Rev. Mod. Phys. \textbf{77}, 633 (2005).}

\bibitem{r1994}  R. Jozsa, \href{https://www.tandfonline.com/doi/abs/10.1080/09500349414552171}{J. Mod. Opt., \textbf{41}, 2315-2323, (1994)}.

\bibitem{g2007}  G. Jaeger, \emph{Quantum information: An overview}, (Springer New York), (2007).

\bibitem{kpa1998} K. {\.Z}yczkowski, P. Horodecki, A. Sanpera, M. Lewenstein, \href{https://journals.aps.org/pra/abstract/10.1103/PhysRevA.58.883}{Phys. Rev.  A \textbf{58}, 883–892 (1998)}; J. Eisert, Entanglement in quantum information theory (Thesis).\href{arXiv:quant-ph/0610253}{ University of Potsdam (2001)}; G. Vidal, R. F. Werner, \href{https://journals.aps.org/pra/abstract/10.1103/PhysRevA.65.032314}{Phys. Rev. A \textbf{65}, 032314 (2002)}.

\bibitem{m2005} M. B. Plenio, \href{10.1103/PhysRevLett.95.090503}{Phys. Rev. Lett. \textbf{95}, 090503 (2005)}.

\bibitem{eak2007} E. Brion, A. S. Mouritzen, and K. M{\o}lmer, \href{https://journals.aps.org/pra/abstract/10.1103/PhysRevA.76.022334}{Phys. Rev. A \textbf{76}, 022334 (2007)}.

\bibitem{idi2010} I. I. Ryabtsev, D. B. Tretyakov, I. I. Beterov, and V. M. Entin, \href{https://journals.aps.org/prl/abstract/10.1103/PhysRevLett.104.073003}{Phys. Rev. Lett. \textbf{104}, 073003 (2010)}.

\bibitem{shd2014}S. Ravets, H. Labuhn, D. Barredo, L. Beguin, T. Lahaye, and A. Browaeys, \href{https://www.nature.com/articles/nphys3119}{Nat. Phys. \textbf{10}, 914 (2014)}.

\end{thebibliography}
\end{document}